\documentclass[aps,prl,twocolumn,superscriptaddress]{revtex4-1}

\usepackage{graphicx}
\usepackage{dcolumn}
\usepackage{bm}

\usepackage{amssymb}

\begin{document}


\title{Transport properties of Andreev polarons \\ in superconductor-semiconductor-superconductor junction \\ with superlattice structure}



\author{Ryotaro Inoue}
  \email{ryoinoue@rs.kagu.tus.ac.jp}
\affiliation{Dept. of Applied Physics, Tokyo Univ. of Science, Tokyo 162-8601, Japan.}%
\affiliation{CREST, Japan Science and Technology Agency, Kawaguchi 332-0012, Japan.}%
  
\author{Kenta Muranaga} 
\affiliation{Dept. of Applied Physics, Tokyo Univ. of Science, Tokyo 162-8601, Japan.}%

\author{Hideaki Takayanagi}
\affiliation{Dept. of Applied Physics, Tokyo Univ. of Science, Tokyo 162-8601, Japan.}%
\affiliation{CREST, Japan Science and Technology Agency, Kawaguchi 332-0012, Japan.}%
\affiliation{MANA, National Inst. for Materials Science, Tsukuba 305-0044, Japan.}%


\author{Eiichi Hanamura}
\affiliation{Japan Science and Technology Agency, Kawaguchi 332-0012, Japan.}

\author{Masafumi Jo}
\affiliation{Research Inst. for Electronic Science, Hokkaido Univ., Sapporo 060-8628, Japan.}%

\author{Tatsushi Akazaki}
\affiliation{NTT Basic Research Lab., Kanagawa 243-0198, Japan.}%
\affiliation{CREST, Japan Science and Technology Agency, Kawaguchi 332-0012, Japan.}%

\author{Ikuo Suemune}
\affiliation{Research Inst. for Electronic Science, Hokkaido Univ., Sapporo 060-8628, Japan.}%
\affiliation{CREST, Japan Science and Technology Agency, Kawaguchi 332-0012, Japan.}%

\date{\today}

\begin{abstract}
Transport properties of a superconductor-semiconductor-superconductor (S-Sm-S) junction with superlattice structure are investigated. 
Differential resistance as a function of voltage shows oscillatory behavior under the irradiation of radio-frequency (RF) waves with the specific frequency of 1.77 GHz regardless of the superconducting materials and the junction geometries.
Experimental data are quantitatively explained in terms of the coupling of superconducting quasiparticles with long-wavelength acoustic phonons indirectly excited by the RF waves.
We propose that the strong coupling causes the formation of novel composite particles, Andreev polarons.
\end{abstract}


\maketitle




Transport properties of Superconductor-Semiconductor-Superconductor (S-Sm-S) junctions have been studied extensively from both experimental and theoretical sides. 
Even in the S-Sm-S junctions where the supercurrent does not flow, diverse phenomena found in the dynamics of correlated quasiparticles, such as multiple Andreev reflections, attract research interests \cite{OTBK,Flensberg}. 
In this paper, we report novel composite particles, which we call Andreev polarons, formed by the strong coupling of quasiparticles with long-wavelength phonons that are indirectly excited by radio-frequency (RF) waves. 
Polarons have been recognized so far as the composite particles, which are formed in the normal conducting (not superconducting) materials or dielectric materials by the electron-lattice interaction such as the deformation potential due to the acoustic phonons and Coulomb interaction in ionic crystals \cite{Polarons&Excitons}.
In conventional superconducting materials, by contrast, it has been unknown that carriers other than Cooper pairs can couple with phonons although the phonon-mediated attractive interaction between electrons is famous for the origin of Cooper pairs.


%


\begin{figure}[htb]
\includegraphics[width=0.8\linewidth, clip]{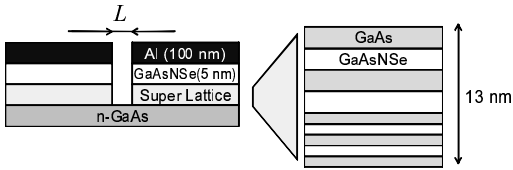}
\caption{Schematic cross-sectional view of the samples. The superlattice structure consists of two GaAs(2 nm)/GaAsNSe(2 nm) layers and three GaAs(1 nm)/GaAsNSe(1 nm) layers.}
\label{fig:Sample}
\end{figure}

Figure \ref{fig:Sample} shows the schematic cross-sectional view of our samples.
GaAs/GaAsNSe superlattice (SL) structures are sandwiched between an n-GaAs layer and two Al superconducting electrodes that form an S-Sm-S junction.
Each of the SL structure consists of two sets of GaAs(2 nm)/GaAsNSe(2 nm) layers and three sets of GaAs(1 nm)/GaAsNSe(1 nm) layers \cite{Uesugi}.
The carrier density and carrier mobility in the n-GaAs layer at room temperature are 5$\times$10$^{18}$ cm$^{-3}$ and 1000 cm$^2$/Vs, respectively.
We prepare samples with different lengths of the n-GaAs layer ($L$), which we identify as the length of the slit between the two Al electrodes (0.5 $\mu$m and 1.0 $\mu$m), while the junction width ($W$) is speculated to be somewhat shorter than the designed value of 0.6 $\mu$m.
We measure the differential resistance of the S-Sm-S junction by means of standard lock-in technique at the temperature of 45 mK under the irradiation of RF waves.
Due to the impedance mismatch at the handmade loop antenna, the irradiated power of RF waves is less than $\sim$1 \% of the monitored RF sweeper power.
The mean free path ($\ell$) and the coherence length ($\xi_N$) in the n-GaAs layer at low temperature of 45 mK are estimated to be $\sim$0.1 $\mu$m and $\sim$0.3 $\mu$m, respectively.


\begin{figure}[htb]
\includegraphics[width=0.7\linewidth, clip]{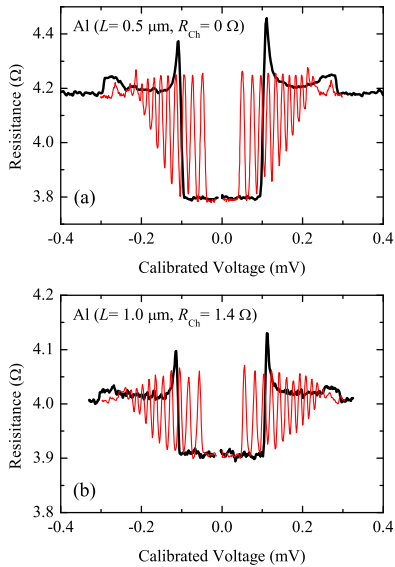}
\caption{Differential resistances as functions of voltage for the length of the n-GaAs layer $L=$0.5 $\mu$m and 1.0 $\mu$m. The heavy black and thin red lines represent the data without RF and with RF irradiation, respectively (Frequency, 1.77 GHz; Power, -10 dBm). Voltage is obtained from the measured voltage ($V_{\rm m}$) and the measured current ($I_{\rm m}$) using the calibration formula $V= V_{\rm m}-R_{\rm Ch}I_{\rm m}$.}
\label{fig:Resistance}
\end{figure}

The measurement results are shown in Fig. \ref{fig:Resistance}.
Although we can see some features in the differential resistance characteristics without RF irradiation, which is closely related to the dynamics of superconducting quasiparticles, the rates of variability of differential resistance in the samples with $L=$0.5 $\mu$m and 1.0 $\mu$m are 10 \% and 2.5 \% , respectively.
These rates of variability are significantly small compared with those in ideal S-Sm-S junctions, and suggest resistive channel be placed in parallel with the junction.
Taking into account the fact that the mean free path ($\ell$) is smaller than the length of n-GaAs layer ($L$), the quasiparticles experience several scatterings during the travel in the n-GaAs layer, and lose the phase information of wave function.   
Especially, in the case where $L$ is longer than $\sim$1 $\mu$m, the channel resistance of the n-GaAs layer is comparable with the interfacial resistance at the Al/SL interfaces, and we cannot neglect the voltage drop across the n-GaAs layer in discussing the quasiparticle dynamics. 
Therefore, when we plot the differential resistance in Fig. \ref{fig:Resistance}, we use the voltage difference between the two Al/SL interfaces obtained by the calibration formula; $V= V_{\rm m}-R_{\rm Ch}I_{\rm m}$.
Here, $V_{\rm m}$ and $I_{\rm m}$ are the measured voltage and current, respectively, and the resistance of the n-GaAs channel ($R_{\rm Ch}$) is estimated to be 0 $\Omega$ and 1.4 $\Omega$ for the sample with $L=$ 0.5 $\mu$m and 1.0 $\mu$m.
Because the resistance of the n-GaAs channel ($R_{\rm Ch}$) is predominantly determined by the additional parallel current channel, the procedure of our calibration can be validated.
\par
On the other hand, the coherence length ($\xi_N$) is comparable with the length of n-GaAs layer ($L$), and the correlation of quasiparticle-pair marginally remains in the n-GaAs layer.
Therefore, we can discuss the behavior of differential resistance characteristics using the Andreev bound state formed in the n-GaAs layer.
The differential resistance characteristics without RF irradiation show peaks at $V$=$\pm V_{\rm peak}$=$\pm$110 $\mu$V in the two samples shown in Fig. \ref{fig:Resistance}.
Although this $V_{\rm peak}$ value is significantly small compared with the twofold superconducting gap voltage of bulk aluminum ($2\Delta_{\rm Al}/e$ = 340 $\mu$V), we regard $V_{\rm peak}$ as the twofold superconducting gap voltage of Al electrodes ($V_{\rm peak} \equiv 2\Delta /e$), because the quasiparticle dynamics drastically change at this value.
The energy of the Andreev bound state under zero voltage is approximately equal to the superconducting gap voltage $\Delta /e$, and the peaks/dips appear in the differential resistance when the superconducting gap edge of either side of Al electrodes coincides with the Andreev bound levels \cite{Aminov}.
\par
Under the irradiation of RF waves (frequency, 1.77 GHz; power, -10 dBm), the differential resistance shows oscillatory behaviors in both of the two samples as shown in Fig. \ref{fig:Resistance}. 
The oscillations are shown only in the voltage near $V_{\rm peak}$= 110 $\mu$V, and the voltage period of the oscillation varies from 25 $\mu$V to 15 $\mu$V, which monotonically decreases with increasing $\vert V \vert$.
We especially note that the minimum voltage period of 15 $\mu$V just coincides with twice the energy of the irradiated RF photons (1.77 GHz $\sim$ 7.3 $\mu$eV).
\par  
The S-Sm-S junctions without SL structures do {\it not} show the oscillatory behaviors in differential resistance under the irradiation of RF waves, while we obtain almost constant {\it resonant} frequency of 1.77 GHz for the junctions with SL structures regardless of the superconducting materials (Al or Nb \cite{InoueJPhys}) and the junction geometries (the length of the n-GaAs layer ($L$) and the junction width ($W$)). 
Therefore, we believe that some phonon modes inherent to the SL structure are excited by the irradiated RF waves and affect the transport properties of the junction.
We will discuss later details of our scenario that the excited phonons strongly couple with the quasiparticles in the S-Sm-S junction and form novel composite particles which we call Andreev polarons.
 
\begin{figure}[htb]
\includegraphics[width=0.7\linewidth, clip]{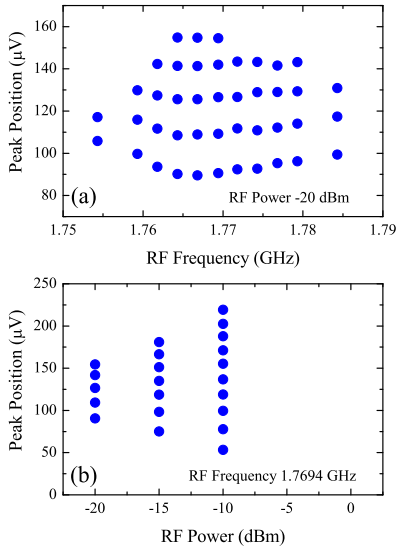}
\caption{Peak position in differential resistances of the sample with $L$=0.5 $\mu$m under the irradiation of RF waves. (a) RF power = -20 dBm, (b) RF frequency = 1.7694 GHz.}
\label{fig:RFPeak}
\end{figure}

In Fig. \ref{fig:RFPeak}, we plot the peak position of the differential resistance as functions of RF wave frequency/power for the sample with $L$=0.5 $\mu$m.
When we slightly change the frequency of the irradiated RF waves, the peaks of the differential resistance shift together maintaining the voltage period almost constant, and take their minima at the {\it resonant} frequency of $\sim$1.77 GHz.
The number of the observed peaks is found to decrease as the RF frequency deviates from the {\it resonant} frequency, and the oscillatory behavior totally disappears when the frequency deviation reaches $\sim$20 MHz.
The number of the peaks is also changed by RF power.
With increasing RF power, new peaks appear in both higher and lower voltage regions while the voltage period of the peaks is almost constant.
\par


First we discuss the involved phonon modes.
Only acoustic modes admit such low-energy phonons that RF excitation is possible. 
Using the sound velocities of acoustic phonons in bulk GaAs (4500 m/s and 2800 m/s for longitudinal and transverse mode, respectively, in (001) direction \cite{GaAsPhonon}), the wavelengths are estimated to be 1.6 $\mu$m and 2.6 $\mu$m for longitudinal and transverse mode.
These wavelengths are comparable with the length of the n-GaAs layer ($L$), and we consider that the standing wave of acoustic phonon mode is excited between two Al electrodes by the RF irradiation.
However, the acoustic phonon modes have poor coupling efficiencies with electromagnetic waves because the electric dipole moments are not induced in the crystals by the acoustic vibrations.
Therefore, this excitation needs another phonon mode which mediates the RF photon energy to the acoustic phonon mode. 
This mediating phonon mode is required to be RF active and inherent to the SL structure.
The most probable candidate is the interfacial phonon mode which runs along the GaAs/GaAsNSe interfaces in the SL structure.
Due to the elastic boundary conditions imposed at the GaAs/GaAsNSe interfaces together with the piezoelectricity in GaAs crystals, the interfacial mode can be RF active and have the {\it resonant} frequency of 1.77 GHz, which is determined by the sizes of GaAs/GaAsNSe interfaces.
Although the dispersion relations of the interfacial phonon modes are nonlinear in general, the sound velocities in low-energy long-wavelength limit are slightly smaller than the transverse sound velocity in bulk GaAs layers (i.e. 2800 m/s in (001) direction).
Thus, the interfacial phonon mode can couple strongly with the acoustic phonon modes of the bulk n-GaAs layer because the phonon energies are close and the pseudo- phase matching condition can be satisfied over a certain distance.
Our speculation of the indirect excitation by the mediating interfacial phonon modes is also supported by the RF frequency dependence of the peak position shown in Fig. \ref{fig:RFPeak}(a).
The slight decrease of peak positions in the vicinity of the {\it resonant} frequency suggests the frequency entrainment with a lower-energy phonon mode caused by nonlinear phonon-phonon interaction. 
\par
As a result, the superconducting quasiparticles coexist with RF-excited long-wavelength acoustic phonons between two Al electrodes.
Taking the coupling with the SL structures into account, both the standing wave of the quasiparticle wave function and that of the phonon vibration are considered to be limited to the thin surface region in the n-GaAs layer, and the quasiparticle-phonon interaction is greatly enhanced due to the dimensional confinement \cite{Hess,Yu,Litovchenko}. 
For the formation of the novel composite particles, only the coexistence and the strong quasiparticle-phonon interaction are essential, and the generating mechanism of Andreev polarons is analogous to that of standard polarons under the electron-phonon interaction \cite{Toyozawa}, which we outlined briefly in the following.
\par
We consider a one-dimensional system of superconducting quasiparticles strongly coupled with a specific acoustic phonon mode vibration.
For the sake of simplicity, we discuss the transport properties of the junction in voltage state using the eigenstates, which exist only under zero voltage in a rigorous sense.
Assuming that quasiparticles have discrete energy levels sufficiently separated from each other, we apply Born-Oppenheimer adiabatic approximation to the energy eigenvalue problem of this coupled system, and obtain the following two equations:
\begin{equation}
\left[
  \left(
  \begin{array}{rr}
    {\hat H} & {\hat \Delta} \\
    {\hat \Delta}^\dagger & -{\hat H}
  \end{array}
  \right)
  - {\tilde V}(x, Q)
\right] 
\left(
\begin{array}{c}
  u \\
  v 
\end{array}
\right)
= \epsilon (Q)
\left(
\begin{array}{c}
  u \\
  v 
\end{array}
\right) ,
\label{eqn:BdG}
\end{equation}
\begin{equation}
\left[
\frac{-\hbar ^2}{2M}  \frac{\partial ^2}{\partial Q^2} + \frac{M\Omega ^2}{2} Q^2 + \epsilon (Q)
\right]
\varphi (Q) = E \: \varphi (Q).
\label{eqn:Phonon}
\end{equation}
\noindent
Equation (\ref{eqn:BdG}) is Bogoliubov-de Gennes equation with quasiparticle-lattice interaction term ${\tilde V}(x, Q)$.
For fixed value of the phonon coordinate ($Q$), we can solve Eq. (\ref{eqn:BdG}) in principle, and obtain the discretized energy eigenvalue of the quasiparticle system ($\epsilon (Q)$) together with the wave functions of quasiparticle eigenstates ($u=u(x;Q)$ and $v=v(x;Q)$). 
Although we predominantly apply the boundary condition of Andreev reflection at the two Al/SL interfaces, normal reflection is also admissible due to the interface barrier.
Equation (\ref{eqn:Phonon}) is Schr\"odinger equation for the wave function representing phonon vibration ($\varphi (Q)$), where harmonic potential is assumed and the energy of the quasiparticle system ($\epsilon (Q)$) is included in the left hand side.
\par
We expand the quasiparticle-lattice interaction term ${\tilde V}(x, Q)$, together with $\epsilon (Q)$, $u(x;Q)$ and $v(x;Q)$, into the power series of normalized phonon coordinate $Q/\Lambda$ where $\Lambda= a(\hbar /M\Omega)^{1/2}$ is the characteristic length which corresponds to the amplitude of the zero-point vibration in the phonon mode with numerical multiple $a \gtrsim 10$; i.e. ${\tilde V}(x, Q) \equiv {\tilde V}_1 (x) (Q/\Lambda) +\cdots ,\quad \epsilon (Q) \equiv \epsilon_0 + \epsilon_1 (Q/\Lambda) +\cdots ,\quad (u, v) \equiv (u_0, v_0) + (u_1, v_1) (Q/\Lambda) +\cdots$, respectively.
It is noteworthy that the zeroth-order term of the quasiparticle-lattice interaction is omitted (${\tilde V}_0(x) \equiv 0$) because it has been already included in ${\hat H}$ and ${\hat \Delta}$ in Eq. (\ref{eqn:BdG}). 
Using the standard perturbation technique, we obtain the energy of the quasiparticle system as $\epsilon (Q) \approx \epsilon_0 - \langle {\tilde V}_1 \rangle Q/\Lambda + O(Q^2/\Lambda^2)$.
Here we note that ${\tilde V}(x, Q)$ is a 2$\times$2 matrix in general, and define the expectation value $\langle {\tilde V}_1 \rangle$ as $\langle {\tilde V}_1 \rangle \equiv \int dx \; (u_0^\ast, v_0^\ast) {\tilde V}_1(x) (u_0, v_0) ^{\mathsf{T}} $. 
We can diagonalize the Hamiltonian in Eq. (\ref{eqn:Phonon}) with $\epsilon (Q) \approx \epsilon_0 - \langle {\tilde V}_1 \rangle Q/\Lambda$ by putting $Q' \equiv Q - \langle {\tilde V}_1 \rangle / M\Omega^2 \Lambda$, and obtain the energy eigenvalue of the total system i.e. the energy of Andreev polaron as 
\begin{equation}
E_n = \epsilon_0 - \frac{\langle {\tilde V}_1 \rangle ^2}{2\hbar \Omega} + \hbar \Omega \left( n+\frac{1}{2} \right) .
\label{eqn:TotalE}
\end{equation}
\noindent
Here the quantum number $n$ can be considered as the number of the phonons which one quasiparticle is dressed with. 

\begin{figure}[htb]
\includegraphics[width=0.6\linewidth, clip]{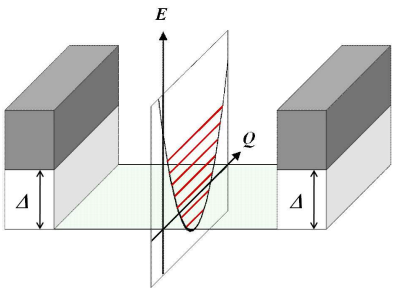}
\caption{Schematic diagram of Andreev polarons together with the adiabatic potential (parabolic line). The superconducting gaps ($\Delta$) in the Al electrodes are also shown.}
\label{fig:APolaron}
\end{figure}

In Fig. \ref{fig:APolaron}, we show the schematic energy diagram of Andreev polarons under zero voltage together with the adiabatic potential ($M\Omega^2 Q^2/2 + \epsilon (Q)$) which the phonon mode vibration is subjected to.
Due to the quasiparticle-lattice interaction, the adiabatic potential takes its minimum at a finite phonon coordinate of $\langle Q \rangle \equiv \langle {\tilde V}_1 \rangle / M\Omega^2 \Lambda$, which is lower than the bare energy of quasiparticles ($\epsilon_0$) by $\langle {\tilde V}_1 \rangle ^2 /2 \hbar \Omega$.
The energy-level spacing of Andreev polarons is approximately equal to that of the involved phonon $\hbar \Omega$ to an accuracy of $O(Q/\Lambda)$.
\par
In order to describe the transport properties of the junction under the voltage state, we have to obtain the occupation probabilities of the involved Andreev bound states, which are different from the equilibrium Fermi distributions in general, and calculated by Boltzmann equation self-consistently, for example \cite{OTBK, Flensberg}.
However, the {\it resonant} voltages where peaks/dips are shown in differential resistance characteristics can be obtained as
\begin{equation} 
eV = 2E_n \equiv 2 \left( \epsilon_0 - \frac{\langle {\tilde V}_1 \rangle ^2}{2\hbar \Omega} + \frac{\hbar \Omega}{2} \right) + 2n \cdot \hbar \Omega . 
\label{eqn:resV}
\end{equation}
\noindent
Thus, with each voltage period of $\delta V = 2\hbar \Omega /e$, the differential resistance characteristics of the junction is modulated because of the novel {\it resonant} condition.
We can state that the new channel concerned with the $n$-th Andreev polaron level participates to the transport of the junction with each voltage period of $\delta V$.
\par
Because the stabilization energy of Andreev polaron is larger than the bare energy of the quasiparticle $\epsilon_0$ ($\sim \Delta$) and the zero-point energy of the excited acoustic phonon $\hbar \Omega /2$, the first term in the right hand side of Eq. (\ref{eqn:resV}) is negative.
From the peak position with the largest $\vert V \vert$, which corresponds to $2E_0/e$, we estimate the magnitude of quasiparticle-lattice interaction $\langle V_1 \rangle \langle Q \rangle/ \Lambda$ as large as $\sim$0.4 meV.
When the irradiated RF power is increased, the expectation value of the phonon coordinate $\langle Q \rangle$ is increased, which is consistent with the appearance of new peaks shown in higher and lower voltage regions (Fig. \ref{fig:RFPeak}(b)).


In conclusion, we investigate transport properties of a superconductor-semiconductor-superconductor (S-Sm-S) junction with superlattice structure.
The differential resistance is measured as a function of bias voltage which is found to oscillate under the irradiation of radio-frequency (RF) waves with a specific frequency of 1.77 GHz.
This oscillatory behavior is quantitatively explained by the strong coupling of superconducting quasiparticles with  long-wavelength acoustic phonon mode vibration which is indirectly excited by RF waves.
We propose the resultant formation of novel composite particles, Andreev polarons.



\begin{acknowledgments}
We thank Dr. D. Tsuya and Dr. E. Watanabe in National Institute for Materials Science (NIMS) Nanotechnology Innovation Center for nanofabrication.
\end{acknowledgments}


\end{document}